\documentclass[english]{article}
\usepackage[T1]{fontenc}

\usepackage[utf8]{inputenc}
\usepackage[english]{babel}
\usepackage{geometry}
\PassOptionsToPackage{hyphens}{url}\usepackage[hidelinks]{hyperref}
\usepackage[normalem]{ulem}
\usepackage{subcaption}
\usepackage{float}
\usepackage{amsmath}
\usepackage{graphicx}
\usepackage{lipsum}
\usepackage[authoryear]{natbib}
\usepackage{xcolor}
\usepackage{lineno}

\makeatletter

\floatstyle{ruled}
\newfloat{algorithm}{tbp}{loa}
\providecommand{\algorithmname}{Algorithm}
\floatname{algorithm}{\protect\algorithmname}
\usepackage{lineno}
\newcommand{\lyxaddress}[1]{
	\par {\raggedright #1
	\vspace{1.4em}
	\noindent\par}
}

\makeatother

\usepackage{babel}
\begin{document}

\title{On the curlometer measurement of field-aligned and perpendicular currents in low Earth orbit: Swarm observations and whole geospace simulations}

\author{R Gajewski$^{1}$\thanks{corresponding author: rafal.gajewski@warwick.ac.uk}, RT Desai$^{1,2}$\thanks{corresponding author: ravindra.desai@warwick.ac.uk}, B Hnat$^{1}$, D Lin$^{3}$, MW Dunlop$^{4}$, \\ M Fillion$^{5}$, G Hulot$^{5}$, Shreedevi P R$^{6}$, M-T Walach$^{7}$, E Panov$^{8}$, J-M Leger$^{9}$, \\ T Jager$^{9}$, D Fischer$^{8}$, W Magnes$^{8}$, JA Blake$^{2}$, T Etchells$^{10}$} 

 \date{}
 \maketitle
 
\vspace{-2em}
\lyxaddress{\centering
$^1$Centre for Fusion, Space \& Astrophysics, University of Warwick, Coventry, UK
}
\vspace{-1.25em}
\lyxaddress{\centering
$^2$Centre for Space Domain Awareness, University of Warwick, Coventry, UK
}
\vspace{-2em}
\lyxaddress{\begin{center}
$^3$Department of Physics, Clemson University, Clemson, South Carolina, USA
\par\end{center}
}

\vspace{-2em}
\lyxaddress{\centering
$^4$Rutherford Appleton Laboratory Space, Science and Technology Facilities Council, Didcot, UK
}
\vspace{-2em}
\lyxaddress{\begin{center}
$^5$Université Paris Cité, Institut de physique du globe de Paris, CNRS, F-75005 Paris, France
\par\end{center}
}
\vspace{-2.8em}
\lyxaddress{\begin{center}
$^6$Institute for Space-Earth Environmental Research, Nagoya University, Nagoya, Japan
\par\end{center}
}
\vspace{-2.8em}
\lyxaddress{\begin{center}
$^7$School of Physics and Astronomy, Lancaster University, Lancaster, UK
\par\end{center}
}
\vspace{-2em}
\lyxaddress{\centering
$^8$Commissariat à l’énergie atomique et aux énergies alternatives, Laboratoire d’électronique et de technologie de l’information, Grenoble, France
}
\vspace{-2em}
\lyxaddress{\begin{center}
$^9$Space Research Institute, Austrian Academy of Sciences, Graz, Austria
\par\end{center}}
\vspace{-2em}
\lyxaddress{\centering
$^{10}$Open Cosmos Ltd, Harwell, UK
}

\begin{abstract}
Measuring field-aligned currents (FACs) using magnetic field observations provides a powerful means to probe the multi-scale interactions between the magnetosphere, ionosphere and thermosphere.
In this study, we benchmark the curlometer technique using
 Swarm spacecraft observations and simulations of the coupled magnetosphere-ionosphere system.
We begin by correlating current density curlometer estimates derived from Swarm tetrahedra with varying spatial scales and bary\-centre locations. This confirms an apparent departure from stationarity for FACs at spatio-temporal scales below 100~km where measurements appear highly uncorrelated.
We then analyse simulated magnetic perturbations, where four-point measurements can always be constructed.
This shows how, even at meso-scales of hundreds of kilometres, time-shifted FAC estimates can diverge significantly from this ground truth.
In both observational and simulated data we find poor tetrahedral configurations can produce spurious perpendicular currents due to numerical instability in the inversion process. This can be mitigated using appropriate quality metrics and high-quality FAC reconstructions still achieved with a tetrahedral face well-aligned to the local magnetic field.
These results highlight the dynamic nature of FACs at large as well as small scales, and underscore the substantial advantages of true four-point observations for their accurate analysis.
\end{abstract}

\subsection*{Key Points}
\begin{itemize}
    \item Swarm curlometer estimates of field-aligned currents appear highly non-stationary below spatio-temporal scales of $\approx$100~km.
    \item Irregular tetrahedral configurations can produce artificial perpendicular currents due to numerical instability in the inversion process.
    \item Testing the time-stationarity assumption with the curlometer applied to simulated data highlights the advantage of four-point measurements.
\end{itemize}

\section{Introduction}
Field-aligned currents (FACs), or \citet{birkeland1908aurora} currents, provide the electrodynamic coupling between the magnetosphere and ionosphere by enforcing current continuity ($\nabla \cdot \mathbf{J} = 0$) along geomagnetic field lines where perpendicular conductivity is low. They arise from magnetospheric convection and pressure gradients, mapping magnetospheric electric fields into the ionosphere and closing via Pedersen and Hall currents \citep{milan2017coupling}. Through Amp\-\`ere’s law ($\mathbf{J} = 1/\mu_0 \nabla \times \mathbf{B}$), FACs perturb the geomagnetic field and redistribute magnetic tension and pressure, thereby modifying field line topology and plasma convection patterns. This feedback regulates energy and momentum transfer from the solar wind, fundamentally shaping both magnetospheric and upper atmosphere structure and dynamics \citep{ebihara2024region1,potemra1985birkeland}.

FACs exhibit a multiscale structure spanning global to kinetic regimes, becoming increasingly intense and dynamic with decreasing spatial scales \citep{McGranaghan25}. At macro-scales ($>1,000$~km), Region 1 and Region 2 current systems, associated with the magnetopause and ring current respectively, close in the ionosphere through Pedersen currents forming a ``Pedersen Loop" that regulates large-scale magnetosphere–iono\-sphere coupling and convection \citep{Untiedt1993}. At meso-scales (${\approx} 100$ km), the spatial variability of the current structure is associated with auroral arcs and current filamentation driven by velocity shears and Alfv\'en waves \citep{lysak2023kaw,tian2026alfven} that may fundamentally build the larger macro-scale structure \citep{Birn2013,Liu2015,Panov2019}. At micro-scales ($<10$ km), FACs become highly filamentary and are associated with Alfvén wave reflections and localised particle precipitation \citep{hallinan2001auroral,zhou2025kilometer}. This multiscale transfer of energy and momentum across disparate spatial regimes, produces fine structure in auroral emissions and intense localised electrodynamic feedback impacting the global system.

Studies of FACs show strong multi-scale spatial and temporal variability across satellite observations, which are used as indications of their underlying stationarity. These are summarised as follows: \citet{ishii1992fac} used Dynamic Explorer 2 to find that stationary FACs did not manifest at scales smaller than 30--75~km. Using the ST5 configuration, \citet{gjerloev2011fac} found that night-side FACs $>$100~km are quasi-stationary, while day-side stability occurs only above 200~km; similarly, \citet{luhr2015fac} showed using Swarm satellites that small-scale FACs (tens of km) are highly dynamic, whereas larger structures ($>$150~km) persist for $>$60~s and exhibit complex topology. \citet{forsyth2017correlations} further used correlation analysis with multiple spacecraft on similar orbits and an additional amplitude-constraint, to demonstrate that larger-scale currents ($>$450~km) appear stationary up to 50\% of the time, whereas small-scale ($<$50~km) currents only 1~\% of the time. A Swarm A–C analysis by \citet{Luhr2025} during the counter-rotation phase extended this by showing that meso\-scale FACs ($>$100 km) remain stable, while small-scale FACs (10–75 km) are only stable over limited ranges and often become non-stationary under enhanced solar wind conditions, likely reflecting transient Alfvénic structures \citep{zhou2025kilometer}.

Despite the extensive use of single- and dual-spacecraft techniques for estimating FACs in low Earth orbit, these approaches rely on assumptions of temporal stationarity, transverse uniformity, or field alignment that become increasingly questionable at meso- and micro-scales \citep{Dunlop2024,FORSYTH2011598}. In regions of rapidly evolving auroral electrodynamics, such assumptions can lead to ambiguity between spatial structure and temporal variability \citep{Luhr2025}. A true multipoint curlometer estimate provides the only direct means of resolving full local current density vector from simultaneous magnetic field gradients, thereby removing ambiguities \citep{DUNLOP1988273}.

The curlometer method is a multi-spacecraft technique used to directly estimate current density by measuring magnetic field gradients across a tetrahedral satellite configuration \citep{middleton2016curlometer}. Originally developed with the four-spacecraft Cluster constellation \citep{Eastwood2008}, it computes the current density via Ampère’s law, using simultaneous magnetic field measurements evaluated at the spacecraft locations. The method has been extended to the Magnetospheric Multi-Scale (MMS) mission, which provides higher-resolution observations, enabling accurate determination of small-scale structures at the magnetopause and magnetotail \citep{Dunlop2024}. With the Swarm constellation, the curlometer has been adapted for two or more satellites in the high field environment of low Earth orbit \citep{dunlop2020swarmcurrents}. This allows estimation of the current through the plane perpendicular to the magnetic field vector via a time-shifted two spacecraft data product, and dedicated full curlometer studies when the spacecraft were separating after launch. 

A key limitation of the Swarm curlometer analysis in low Earth orbit (LEO) is the time-stationarity assumption employed to create four measurement points from three spacecraft. Theoretical studies of the curlometer have tested its performance in idealised current structures, using Taylor-expanded magnetic fields to simulate realistic gradients. These analyses quantified how inter-spacecraft spacing and field curvature affect the accuracy of current density estimates, establishing the method’s limits in both planar and structured magnetospheric current systems \citep{FORSYTH2011598}. Relating this to FAC observations in LEO is more difficult as the current FAC structure is complex and not well constrained. Global magnetospheric simulations coupled to ionosphere and thermosphere models are, however, providing increasingly accurate representations of currents flowing in and out of the ionosphere \citep[e.g.][]{wang2025mage,Desai2021} and have yet to be used as a ground truth for curlometry studies in high field environments. 

In this study, we employ and evaluate the curlometer technique for correlating FAC structure in LEO. In Section 2 we describe the theory of deriving the curlometer from multipoint observations. Section 3 describes the analysis of Swarm data while Section 4 tests the time-stationary assumption inherent in this analysis. Section 5 moves beyond this to analyse four-point data obtained from virtual satellites within numerical simulations of the magnetosphere-ionosphere system. Section 6 then summarises and discusses the results. Previously, the \citet{Ritter2006} study examined the calculation of FACs within magnetospheric simulations from two synthetic spacecraft flying side-by-side. In this study, we build upon previous studies evaluating the applicability of the curlometer technique for high latitude FAC estimates \citep{Luhr2025,dunlop2020swarmcurrents, dunlop2020ionospheric,forsyth2017correlations} to elucidate the advantages of curlometer measurements with four-point measurements. In particular, we quantify the effects of tetrahedron geometry and the time-stationarity assumption on FAC reconstruction accuracy using both Swarm observations and physics-based simulations as ground truth.
\section{The curlometer method}
\newcommand{\unit}[1]{\hat{\mathbf{#1}}}
\newcommand{\dB}[1]{\Delta\mathbf{B}_{#1}}
\newcommand{\dR}[1]{\Delta\mathbf{r}_{#1}}

\begin{figure}
    \centering
    \includegraphics[width=0.4\linewidth]{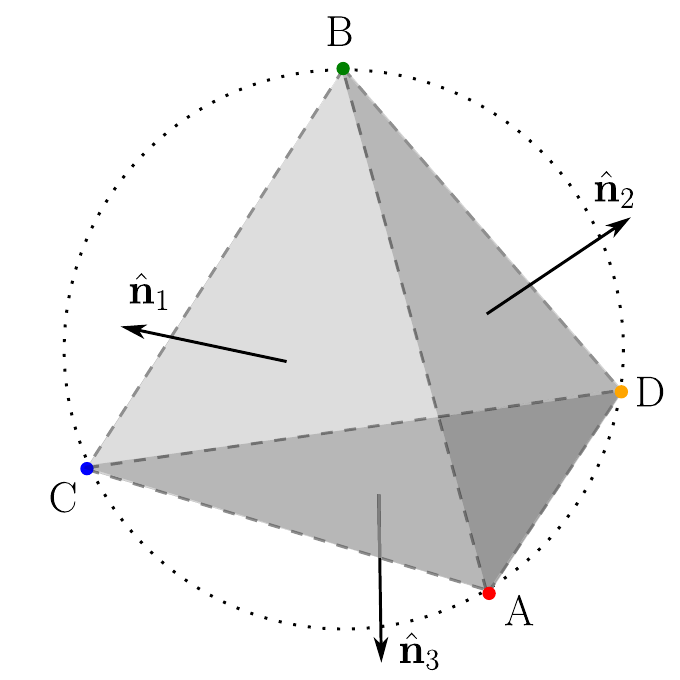}
    \caption{Illustration of the curlometer tetrahedron: Nodes denoted A, B, C and D represent spacecraft positions. In the case of Swarm, measurements from one of the three available spacecraft A, B or C shifted forward or backward in time are used to obtain the fourth tetrahedron node D. Magnetic field measurements taken at the nodes are used to compute average current densities flowing through the tetrahedron faces. An example choice of three faces with vector normals (denoted $\unit{n}_1, \unit{n}_2$ and $\unit{n}_3$) is shown. The dotted circle marks the surface of a sphere of which the interior volume is used to calculate the Robert-Roux regularity parameter.}
    \label{fig:curlometer_illustration}
\end{figure}

Throughout this study, we compute current density vectors using the standard curlometer method, which we summarise in this section (for full details see \cite{Dunlop2024} and \cite{middleton2016curlometer}). The curlometer relies on finite differences in vector magnetic field measurements taken at the four nodes of a 4-point spacecraft constellation (see illustration in Figure \ref{fig:curlometer_illustration}). Specifically, for position difference vectors between the spacecraft $i$ and $j$ denoted $\Delta\mathbf{r}_{ij}$ and the corresponding magnetic field vector differences $\Delta\mathbf{B}_{ij}$, a linear approximation to Amp\`ere’s law integral gives the average current density through a tetrahedron face spanned by nodes $i$, $j$ and $k$,

\begin{equation}
J_{ijk} = \mathbf{J}\cdot\hat{\mathbf{n}}_{ijk} = \frac{1}{\mu_0}\frac{\Delta\mathbf{B}_{ik}\cdot\Delta\mathbf{r}_{jk} - \Delta\mathbf{B}_{jk}\cdot\Delta\mathbf{r}_{ik}}{|\Delta\mathbf{r}_{ik}\times\Delta\mathbf{r}_{jk}|}
\label{eq:curlometer}
\end{equation}
where $\hat{\mathbf{n}}_{ijk}$ denotes the unit vector normal to the face $ijk$ and $\mathbf{J}$ is the average current density vector inside the tetrahedron volume.

A four-point constellation contains enough information to reconstruct vector $\mathbf{J}$, by solving the set of three simultaneous equations \eqref{eq:curlometer} for three different faces of the tetrahedron; expressed in matrix form for faces 1,2 and 3, this can be achieved by matrix inversion,

\begin{equation}
\mathbf{J} = 
\begin{pmatrix}
\unit{n}_1 \\
\unit{n}_2 \\
\unit{n}_3
\end{pmatrix}^{-1}\cdot\begin{pmatrix}
J_1 \\
J_2 \\
J_3
\end{pmatrix}
\label{eq:matrix-inv}
\end{equation}
where $\unit{n}_n$ denotes the square matrix of unit vector components for the three faces $n=1,2,3$ and the $J_n$ denotes the current density through the different faces. Throughout this work we express the components of vector $\mathbf{J}$ in the solar-magnetic (SM) coordinate system.

The quality of the estimate crucially depends on the geometry of the tetrahedron, uncertainties in the magnetic field measurements and non-linearities in the magnetic field \citep{Robert1998Accuracy}. To evaluate the quality of the estimates, we also compute the dimensionless ratio $\text{|div}\mathbf{B}|/|\text{curl}{\mathbf{B}}|$ which under ideal estimation conditions is equal to zero, and intervals with the ratio < 0.3 - 0.5 are typically considered reliable \citep{middleton2016curlometer}. In general, however, the ratio alone is not sufficient to determine the accuracy of the measurement; further considerations of the spatial scale, orientation, and measurement uncertainties in both the observed magnetic field structure and the tetrahedral geometry are required \citep{Robert1998Accuracy}. 

Choosing index $i$ as the reference node, we compute the estimate of the ratio with,

\begin{equation}
\frac{\text{|div}\mathbf{B}|}{|\text{curl}\mathbf{B}|} = \frac{1}{\mu_0|\mathbf{J}|}\left|\frac{\sum_\text{cyclic}\dB{ij}\cdot\dR{ik}\times\dR{il}}{\dR{ij}\cdot\dR{ik}\times\dR{il}}\right|
\end{equation}
where the sum runs over cyclic permutations of $j$, $k$ and $l$ and $|\mathbf{J}|$ denotes the magnitude of our estimate of the average current density vector. 

The inversion process in the curlometer requires a linearly-independent set of estimates through the faces \citep{DUNLOP1988273}. Because of this, accurate three-dimensional reconstruction of the current density demands the spacecraft configuration to resemble a regular tetrahedron. To gauge the dependence of our results on the tetrahedron geometry, we evaluate the regularity using the one-dimensional, normalised Robert-Roux parameter \citep{middleton2016curlometer} defined as the ratio of the tetrahedron volume to the volume of a sphere by which the tetrahedron is bounded. A perfectly regular tetrahedron then has the normalised Robert-Roux parameter equal to 1, as depicted in Figure \ref{fig:curlometer_illustration}. One-dimensional quality parameters provide the simplest indicators of poor tetrahedral geometry, which leads to inaccuracies in the estimated current density. However, they do not identify which specific geometric properties are responsible for this degradation in accuracy. Elongation and planarity are a two-dimensional parameter set that serve a similar purpose while providing greater insight into the tetrahedron shape \citep{1998ISSIR}. These parameters are defined in terms of the eigenvalues of the volumetric tensor constructed from the tetrahedron vertices,
\begin{equation}
    R_{jk} = \sum_{\alpha=1}^N (r_{\alpha j} - b_j)(r_{\alpha k} - b_k)
    \label{eq:vol_tens}
\end{equation}
where $\mathbf{r}_\alpha$ corresponds to the position vector of the tetrahedron vertex $\alpha$ with vector components $j,k=x,y,z$, while vector $\mathbf{b}$ denotes the tetrahedron barycentre. The elongation $E$ and planarity $P$ are then defines as,
\begin{align}
    E = 1 - \frac{b}{a}\\
    P = 1 - \frac{c}{b}
\end{align}
for eigenvalues $a,b$ and $c$ of the volumetric tensor $R_{jk}$ in equation \eqref{eq:vol_tens} such that $a\geq b \geq c$ \citep{Robert1998Tetrahedron}. Defined in this way, the ideal geometry corresponding to a regular tetrahedron is characterised by $E,P =0$. The parameters enable the classification of the tetrahedron shape by its value of $E$ and $P$ with a corresponding accuracy in the current density estimate which is expected for a given geometry. \citet{Robert1998Tetrahedron} identify 5 types of tetrahedron shapes for values of $E$ and $P$ falling in the intermediate range between 0 and 1. For example, shapes with $0\leq E \leq 0.4$ and $0.6 \leq P \leq 1.0$ are classified as a flat `pancake' geometry. A simulation of a tetrahedron crossing an analytical current structure shows that the relative error in the measured current density introduced by geometric distortion reaches >10\% for flat tetrahedra with large $P$. The corresponding Robert-Roux parameter takes on a value in the range <0.1 - 0.2 (see Figures 13.10 in \citet{Robert1998Tetrahedron} and 16.7 in \citet{Robert1998Accuracy})
Nevertheless, a complete characterisation of the geometric influence on the current density estimate also requires analysis of the volumetric tensor eigenvectors, which determine the tetrahedron's orientation. In this work, only the Robert-Roux parameter is shown alongside our results, but elongation and planarity are discussed where it may provide additional insight.

Finally, we note that alternative formulations of the curlometer technique exist, which enable specifying additional constraints in the estimation \citep[e.g.][]{Fillion2021,Hnat2023,Singh2026IntegrationFree}. 

\section{Swarm data analysis}
\begin{figure}[t!]
    \includegraphics[width=\textwidth]{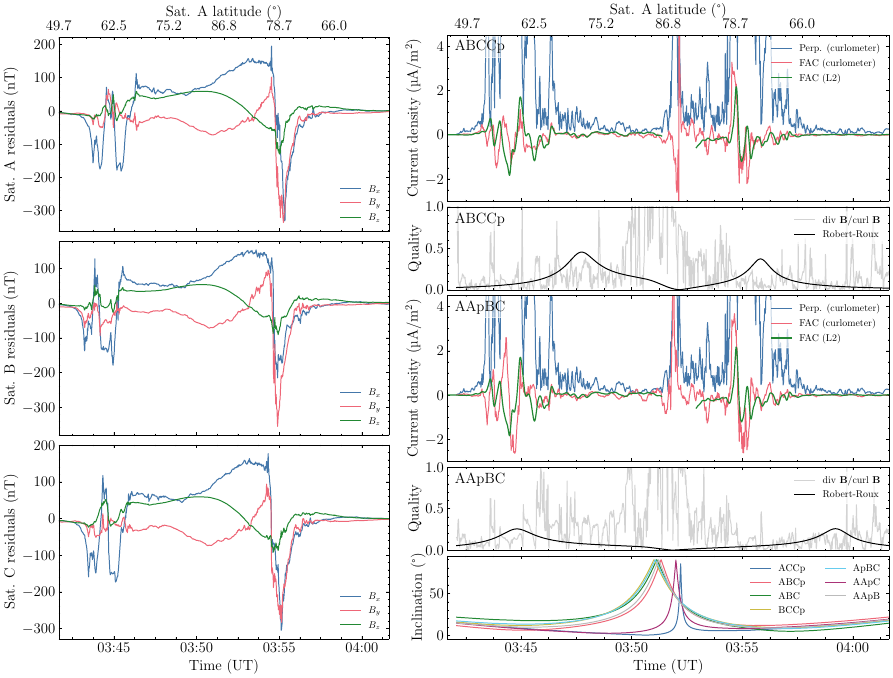}
    \caption{\textbf{Left}: Cartesian components of the magnetic field residuals (SM) for Swarm spacecraft A B and C collected for an auroral passing over the northern hemisphere on 17 June 2014. \textbf{Right}: The corresponding current density estimates obtained using the curlometer decomposed into the field-aligned component and the remaining perpendicular magnitude, for the spacecraft tetrahedron configurations ABCCp (top) and AApBC (middle), where Cp and Ap denote measurements taken with nodes C and A time-shifted backwards by 25 seconds respectively. The L2 FAC data product is shown for comparison. The quality measures show the ratio $\text{|div}\mathbf{B}|/\text{|curl}\mathbf{B}|$ and the Robert-Roux tetrahedron regularity parameter. The bottom plot shows the inclination of the tetrahedron face normals with respect to the field direction. The Kp index (not shown) is 1.00 during this interval.} 
    \label{fig:swarm1}
\end{figure}
We focus our analysis on Swarm trajectories during the period April-June 2014 when the three spacecraft were orbiting in close alignment. Specifically, we use selected auroral crossing events from the Methodology Inter-Comparison Exercise (MICE) study \citep{dunlop2020ionospheric} identified as suitable for calculation of current densities. In Section \ref{sec:swarm_corr}  we use 1 Hz level 1b (version 0701) magnetometer data (\url{https://swarm-diss.eo.esa.int/}) in the calculation of the estimate correlations. 
Prior to evaluating current density estimates, we subtract the internal, ionospheric and magnetospheric contributions to the magnetic field using the predictions of the CHAOS-8.3 model. Tested applications of this step with and without the inclusion of the ionospheric and magnetospheric contributions showed difference of $\sim0.01$ \textmu A/m$^2$ due to the ionosopheric contribution, and an order of magnitude smaller difference due to the magnetospheric field in the curlometer FAC estimates, for this time period. We use the Level 2 (L2; version 0401), FAC data product from spacecraft A and C for comparison in our results \citep{Ritter2013} where both spacecraft are timeshifted to obtain a FAC estimated through the sampled 2D area. In Section \ref{sec:rcm} we use the same Swarm positions and therefore tetrahedral geometry to sample simulated magnetic perturbations.

To apply the curlometer to the measurements taken with Swarm, we use the time-shift method which uses measurements from one of the three spacecraft, shifted forward or backward in time to generate the fourth node in the constellation. This is considered applicable when the underlying current structure varies slowly on the spatio-temporal scale of the tetrahedron volume. 

Figure \ref{fig:swarm1} shows an example of the curlometer application for a 25-second delay, for an auroral passing over the northern hemisphere on 17 June 2014. Here the value of 25 seconds was chosen to maximise the tetrahedron regularity over the auroral regions. We adopt the notation used in \citet{swarm2016curlometer}, using `p' following the label corresponding to one of the spacecraft to denote a shift backwards in time e.g. `Ap' to denote delayed measurements from spacecraft A. The Figure shows the magnetic field residuals for each spacecraft and the corresponding current density estimates for a pass over the northern hemisphere on 17 June 2014 for two tetrahedron configurations ABCCp and AApBC. Period of strong FACs, corresponding to each pass over the auroral oval are present near 03:{4}4 and 03:55, each lasting about 6 minutes and comprising of Region 1 and Region 2 current systems.

The time-shift chosen is the same for both configurations in Figure \ref{fig:swarm1}. While the choice of time-shift provides some level of control over the tetrahedron regularity, the two configurations sample slightly different volumes and their regularity is only partially aligned with the auroal oval; ABCCp (first from the top in the right panel of Figure \ref{fig:swarm1}), and AApBC (third from the top in the right panel of Figure \ref{fig:swarm1}). There is a tetrahedron regularity mismatch between the two configurations, between the first crossing of the auroral oval (around 03:45 UT) and the second cross of the auroral oval (around 03:55 UT). Depending on the shape of the trajectories, the choice of time-shift can sometimes be adjusted independently for the two configurations such that the regularity of the resulting tetrahedron peaks over the auroral region. We focus on isolating the geometric effects on the quality of estimates in Section \ref{sec:rcm} where we make a direct comparison between estimates obtained using a Swarm and a regular tetrahedron.

The corresponding FAC estimates appear substantially different in the first part of the auroral crossing, which could reflect a combination of geometric effects and rapid temporal evolution of the current structure (dayside/nightside differences). However, this interpretation is not clearly borne out when inspecting the ratio $\text{div}\mathbf{B}/\text{curl}\mathbf{B}$; while the values do not appear consistent for the two different spacecraft configurations, averaging the ratio separately over the auroral zone yields about 0.16 in both regions. 
This may also be associated with the angle at which the auroral oval is crossed, since FACs are more stable on the
dawn and dusk sides as compared to the noon-midnight \citep{Lukianova20,ForsythDuskDawn18}. The L2 estimate is also expected to suffer from inaccuracies during active times, as the plane spanned by spacecraft A, C and its delayed counterparts is not exactly perpendicular to the Earth's magnetic field lines. This may explain the discrepancy between curlometer estimates and the L2 FAC visible in the first part of the auroral crossing. 

The irregular configurations of the Swarm tetrahedra in the curlometer give rise to artificially exaggerated magnitudes of perpendicular currents. This is because the curlometer requires a linearly independent set of current density estimates. However, the Swarm trajectories provide irregular coverage of the different directions; the tetrahedron has planarity > 0.9 over both crossings of the auroral oval, characterising the tetrahedron as significantly flattened (`pancake' geometry, see \citet{Robert1998Tetrahedron}).   Accurate estimation of the perpendicular currents would require the tetrahedron faces to point away from the magnetic field. The inclination plot in the bottom-right of Figure \ref{fig:swarm1} shows that the vector normals of the tetrahedron faces are mostly field-aligned over the auroral regions. The inaccuracy enters through the matrix inversion process in Equation \eqref{eq:matrix-inv}, where insufficient coverage in three dimensions makes the matrix near singular, which leads to large matrix inversion terms that amplify small amplitudes and noise. In Section \ref{sec:rcm}, we use simulated data to demonstrate this geometric effect, and show that despite this limitation for perpendicular currents, the sampling of the field direction by Swarm trajectories in this period is sufficient to accurately estimate FACs with the curlometer. In addition, we show that simple changes to the orbits can aid in regularising the tetrahedron geometry and mitigating the limitation for perpendicular currents.  

The curlometer calculation shows a large spike in current near the orbit crossover, due to the extreme geometrical distortion where the inclination of the tetrahedral faces rapidly increase. This is also reflected in $\text{div}\mathbf{B}/\text{curl}\mathbf{B}$, which diverges in the vicinity of the crossover. The L2 product is not provided near the geographic pole where the crossover occurs \citep{Ritter2013}. 

\section{Testing the time-stationarity assumption with Swarm data} \label{sec:swarm_corr}

\begin{figure*}[h!]
    \includegraphics[width=\textwidth]{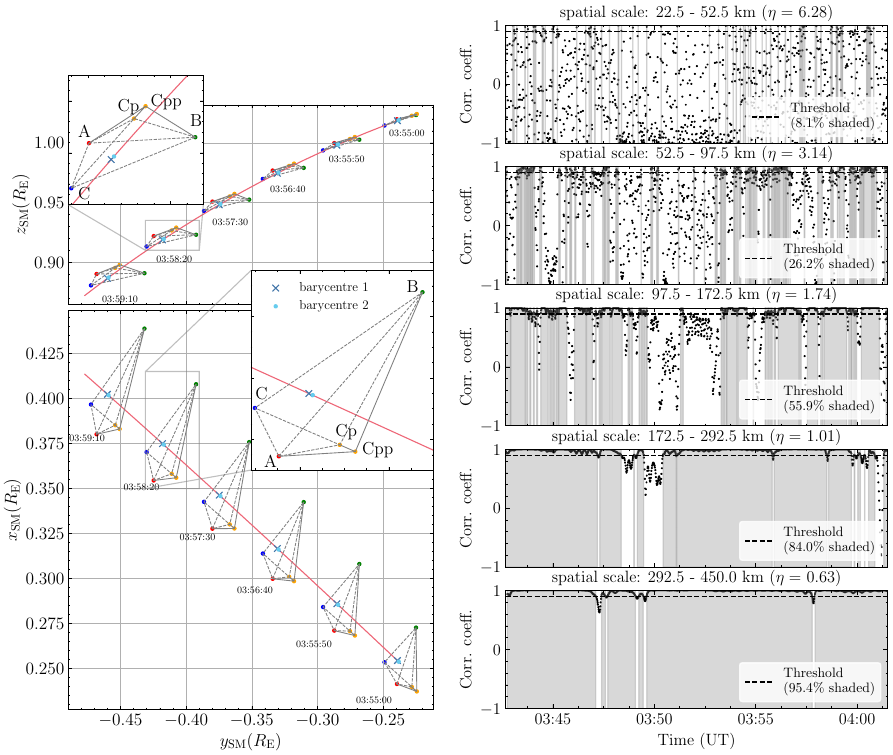}
    \caption{\textbf{Left}: Part of the Swarm trajectory (17 June 2014) used to evaluate correlations between estimates, shown in the $y_\text{SM}$-$z_\text{SM}$ plane (top) and $x_\text{SM}$-$y_\text{SM}$ plane (bottom), expressed in terms of the Earth's radius $R_\text{E}$. Nodes Cp and Cpp denote the positions of spacecraft C delayed by 22 and 26 s respectively. The distance between the barycentre positions of the two tetrahedra in the enlarged view is around 7.5 km. \textbf{Right}: Correlation coefficients computed between FAC estimates of tetrahedron configurations ABCCp and ABCCpp, band-passed at different spatial-scales. The shaded area marks correlation sections above the chosen threshold of 0.9.}
    \label{fig:corr_test1}
\end{figure*}
The choice of the time-shift gives us fine control over the position and time corresponding to the tetrahedron barycentre. This enables testing the validity of the time-stationarity assumption used in the time-shift method \citep{swarm2016curlometer} by comparing estimates corresponding to tetrahedron volumes with small differences. In this section we follow some of the methodology of \citet{forsyth2017correlations} developed for testing the stationarity assumption in single spacecraft analyses.  Specifically, we analyse the correlation coefficient of the curlometer estimates made with the time-shift method, to check whether the time-stationarity assumption is applicable for the curlometer with the typical Swarm trajectories in the period April-June 2014.
\begin{figure*}[t!]
        \includegraphics[width=\textwidth]{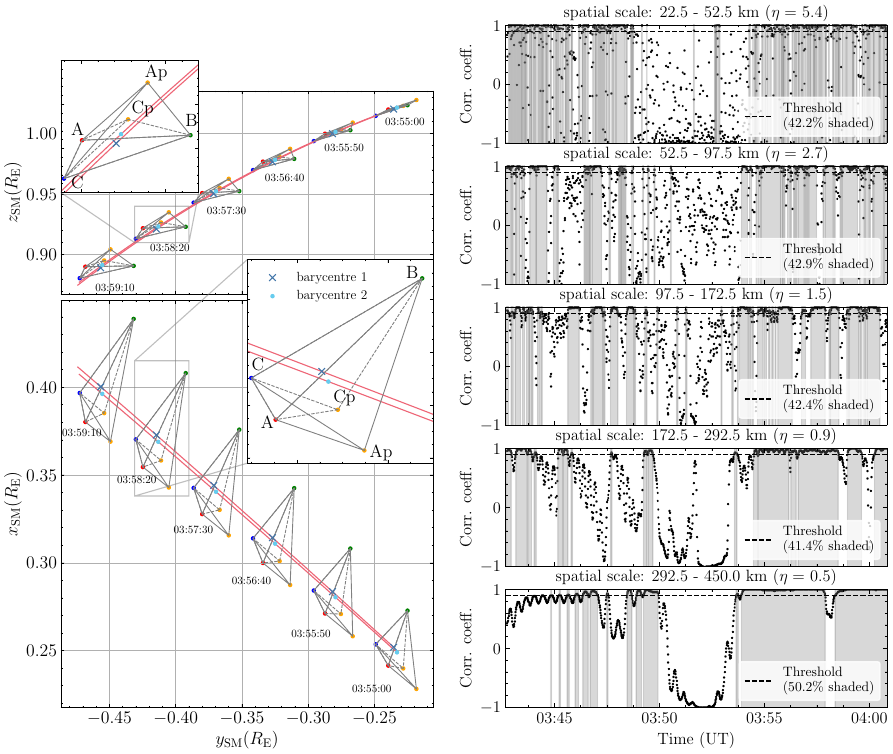}
    \caption{\textbf{Left}: Part of the Swarm trajectory (17 June 2014) used to evaluate correlations between estimates, shown in the $y_\text{SM}$-$z_\text{SM}$ plane (top) and $x_\text{SM}$-$y_\text{SM}$ plane (bottom), expressed in terms of the Earth's radius $R_\text{E}$. Nodes Cp and Ap denote the positions of spacecraft C and A delayed by 22 s respectively. The distance between the barycentre positions of the two tetrahedra in the enlarged view is around 35 km. The \textbf{Right}: Correlation coefficients computed between FAC estimates of tetrahedron configurations ABCCp and AApBC, band-passed at different spatial-scales. The shaded area marks correlation sections above the chosen threshold of 0.9.}
    \label{fig:corr_test2}
\end{figure*}

We band-pass filter the frequency spectra of FAC estimate time traces (such as those shown on the right of Figure \ref{fig:swarm1})  through a Hann window corresponding to different spatial scales, where the spatial scale is calculated as the signal time period multiplied by the spacecraft velocity.  We then analyse the Pearson correlation coefficient for band-passed estimates in two scenarios:

\begin{enumerate}
    \item Estimates made with configuration ABCCp, and the same configuration with the node Cp delayed by a further 4 seconds (denoted ABCCpp). The additional delay results in a slight stretching of the tetrahedron volume, with a small displacement of the barycentre position (around {7.5} km, see Figure \ref{fig:corr_test1}). Due to the small delay difference, the estimates represent averages for different barycentre time. Defining the barycentre time as the average time of the measurements taken by the four nodes, a 4 second delay difference results in a 1-second difference in the barycentre time. The 1 second difference on the barycentre matches the resolution of the Swarm data, such that putting the estimates onto a common time series does not require interpolation.
    \item Estimates made with two different node configurations of a Swarm tetrahedron, ABCCp and AApBC. For typical time-shifts of 22 seconds the barycentre positions are separated by about 35 km (see Figure \ref{fig:corr_test2}). Since both configurations contain a single delayed node, they have the same barycentre time. 
\end{enumerate}

A case study for the first of these tests is shown in Figure \ref{fig:corr_test1} for a pass of the Swarm spacecraft over the northern hemisphere on 17 June 2014. The traces of the correlation coefficient show a clear pattern across spatial scales, being less correlated at smaller scales, in agreement with previous studies \citep{forsyth2017correlations,Luhr2025,gjerloev2011fac,ishii1992fac}. Since the estimates are separated by a time difference much smaller than the delay used in the time-shift method, lack of correlation at small scales points to the fact that the underlying small-scale current structure are evolving too rapidly for a time-shifted Swarm tetrahedron to capture accurately.

The plots in the right-hand panels of Figures \ref{fig:corr_test1} and \ref{fig:corr_test2} also show the mean value of the dimensionless ratio $\eta$ , which we define as the ratio of the mean spacecraft separation and the mean spatial scale of the bandpass filter. The ratio changes as the volume of the tetrahedron evolves during the crossing; here we display the value averaged over the crossing interval. Different values of the $\eta$ parameter across the correlation plots give some indication of spatial scales at which to expect a transition into a non-stationary regime. For $\eta$ of around 1.0 (in bottom two panels of Figure \ref{fig:corr_test1}), a point at which one would expect the breakdown of the stationarity assumption to occur \citep{FORSYTH2011598}, the portion of shaded area corresponding to the correlation coefficient > 0.9 over the presented interval falls below 90\%. Less stringent requirements on the correlated interval yield a higher value for $\eta$; the portion of the shaded area falls below 50\% between panels 2 and 3, corresponding to $\eta$ of about 2.0. Similar analysis repeated over many auroral crossings could provide a generally applicable limit for the transition into the non-stationary regime at the desired correlation thresholds.
 
The results in Figure \ref{fig:corr_test2} show the corresponding case study for the second test scenario, for the two different configurations ABCCp and AApBC. The shaded area indicates that the estimates correlate nearly the same across scales, with a small increase at the largest scale. The two configurations have the same barycentre time, and the left panel of Figure \ref{fig:corr_test2} indicates that they generally also have a large volume overlap, apart from the interval near the orbit cross-over where the volume becomes very small and we do not see good correlations. This could explain why the two configurations are able to capture small-scale current structures coherently. However, this correlation test only measures linear correlation, and does not take into account changes in the amplitude which can be vastly different \citep{forsyth2017correlations}. This can also be seen in the right panel of Figure \ref{fig:swarm1}. We also note, that in both test cases the cross-correlations on the scales lower than the dimensions of the tetrahedron have limited reliability for two reasons. First, is that due to the sampling frequency of the Swarm data used and the size of the band-pass window at smaller scales, each correlation coefficient was calculated using only a few samples. The second is that the curlometer does not accurately resolve current structures below the characteristic spatial scale of the tetrahedron size \citep{FORSYTH2011598}.

\begin{figure}[h!]
    \centering
    \includegraphics[width=\linewidth]{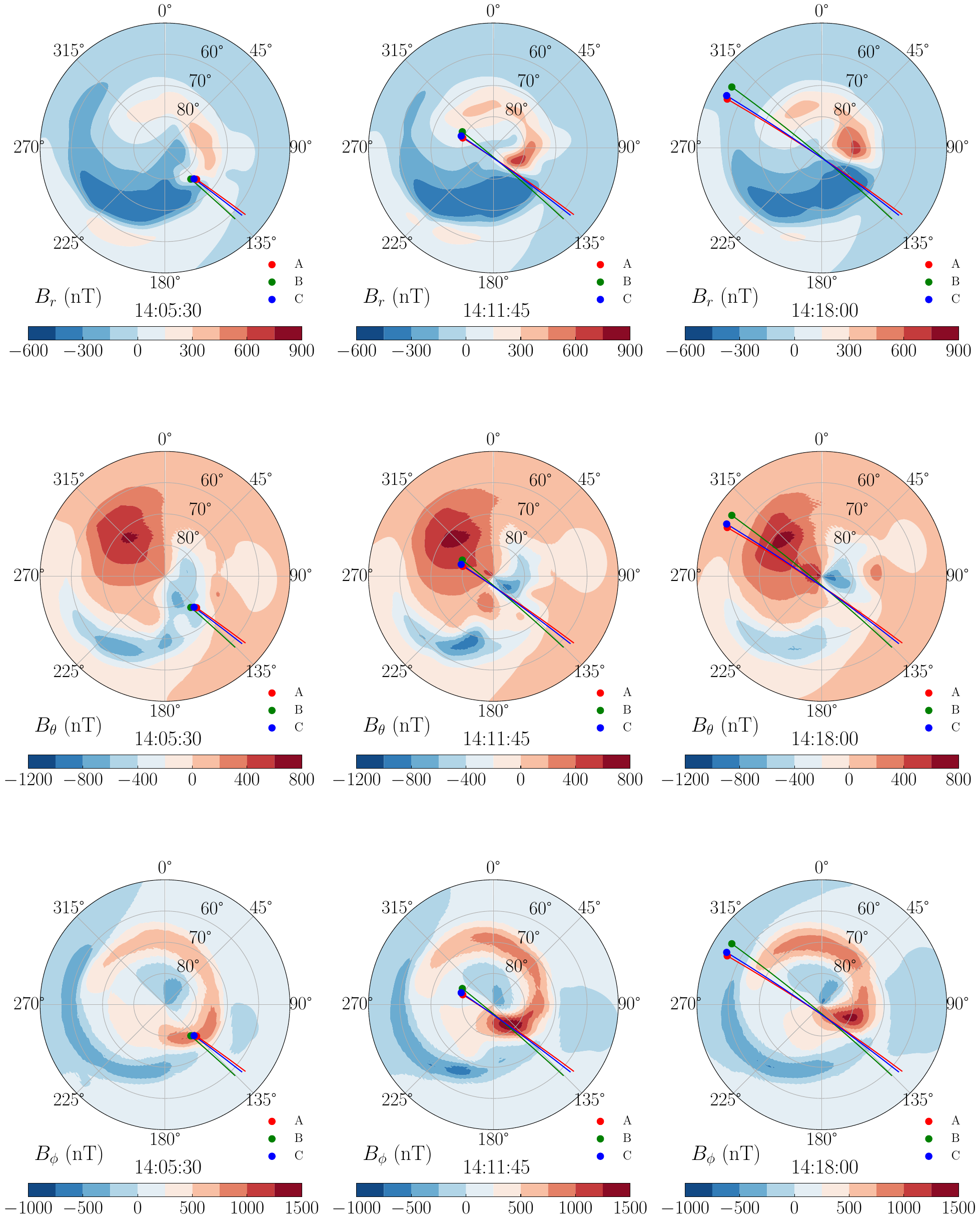}
\caption{Radial, R, meridional, $\theta$, and azimuthal, $\phi$ components of the simulated magnetic field perturbations (on a geographic grid)  corresponding to the conditions on 23 March 2024, {14:05:30–14:18:00 UT,} at 400 km altitude in the northern hemisphere. The plot shows the evolution of the perturbations and {a segment of the 29 May 2014, 13:26--13:46 UT Swarm spacecraft} trajectories used to sample them.}
    \label{fig:polarB}
\end{figure}

\section{Whole Geospace Simulations} \label{sec:rcm}

To further probe the curlometer technique in low Earth orbit, we next consider simulated fields produced from the MAGE (Multiscale Atmosphere-Geospace Environment) model \citep{Merkin2025}. This encompasses the GAMERA magnetohydrodynamic (MHD) model that captures the solar wind-magnetosphere interaction \citep{Zhang2019}, a coupled thin shell ionosphere that provides an electrostatic solution for Hall and Pedersen ionospheric currents on a 2-D grid \citep{Merkin2010} and a ring current model that solves for energy-dependent convection drift of electrons and ions in the inner magnetosphere \citep{Toffoletto2003}. This coupled whole geospace model has the capability to characterise magnetosphere-ionosphere current systems, including current sheets at the bow shock and magnetopause, ring current, field-aligned current, and ionospheric Pedersen and Hall currents. The model can then use Biot-Savart integration to derive magnetic perturbations caused by these currents in different domains in geospace, including inside the inner boundary of the MHD domain for LEO satellite orbits and on the ground \citep{Sorathia2023}.
The magnetic perturbations were calculated on a global geographic grid with 0.5$^\circ$ angular resolution at discrete altitude levels between 300 and 500 km with a 10 km resolution. No additional instrumental noise has been introduced to the simulated perturbations, allowing us to isolate the effects of the tetrahedron geometry and time evolution of the field in the analysis of the estimation process. Using the magnetic perturbations directly also allows us to ignore possible artifacts arising in the process of subtraction of the background field, which has to be done when working with real experimental measurements.

We use an existing simulation corresponding to 23 March 2024, 14:00-15:00 UT, where MAGE data is output with a 5 s resolution. The magnetic perturbations are shown in Figure \ref{fig:polarB} together with the Swarm spacecraft trajectories taken from 29 May 2014 13:31:10--13:4:34:40 when the three spacecraft were closely aligned enabling three-spacecraft curlometer estimates. The magnetic perturbations become stronger across this period particularly within the $\phi$ component, when the virtual Swarm spacecraft trajectories encounter corresponding currents directed into and out of the ionosphere. The simulation covered a geomagnetically quiet interval with a Kp index $<$2.00.

\subsection{Geometrical Effects}

\begin{figure}[h!]
    \centering
    \includegraphics[width=0.85\linewidth]{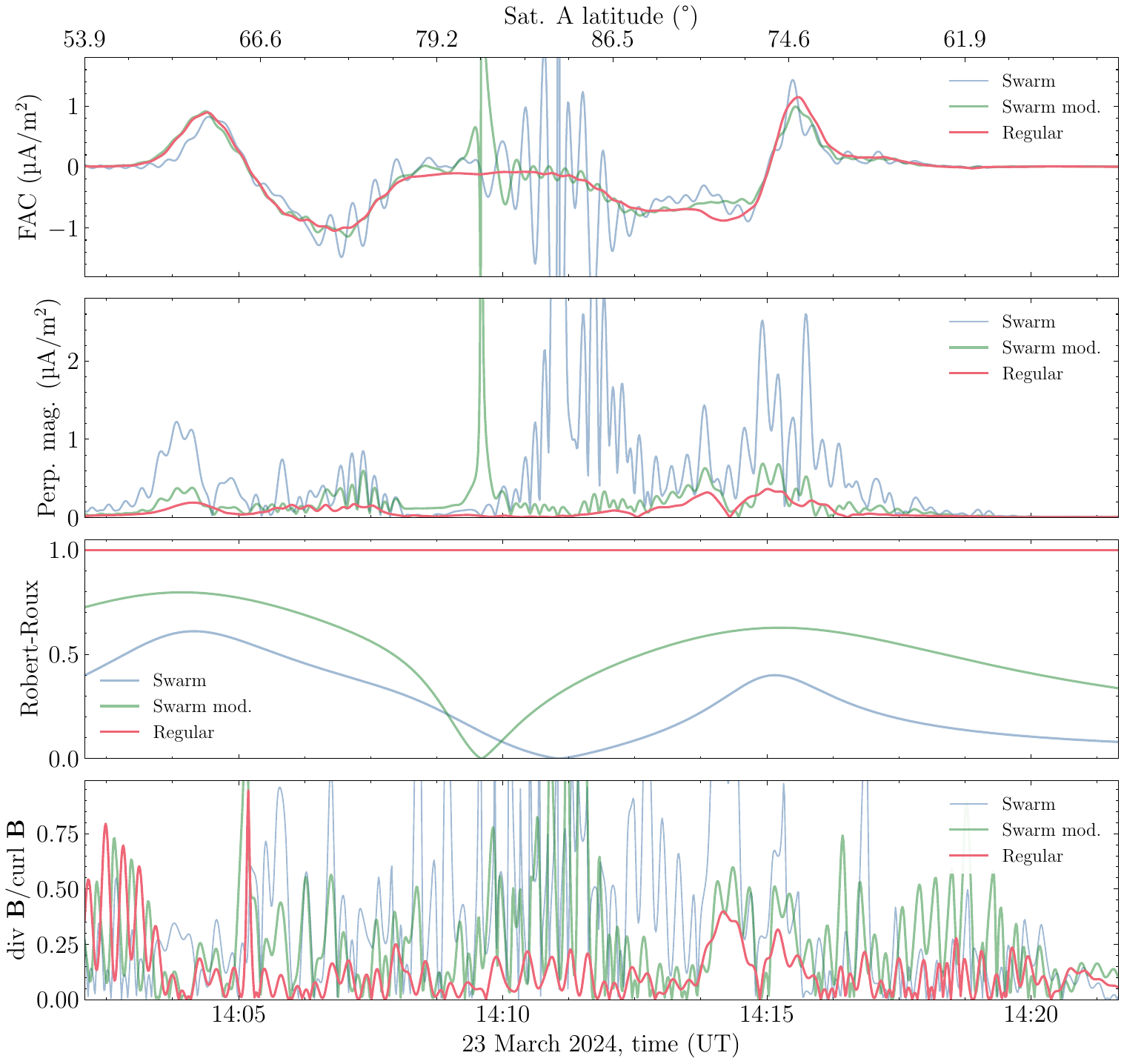}
    \caption{Curlometer estimates of FAC (top)  and the magnitude of the  perpendicular current (second from top) and the corresponding quality measures: Robert-Roux regularity parameter (middle) and the trace of $\text{div}\mathbf{B}/\text{curl}\mathbf{B}$ (bottom) for the 22 April 2014 Swarm trajectory sampling a single time-slice (14:08 UT) of the simulated magnetic field perturbations. The current densities were estimated using the curlometer with the Swarm tetrahedron in the configuration ABCCp with a 25 s time-shift, and a regular tetrahedron with its 4 nodes around the barycentre of the Swarm tetrahedron with an edge-length of 150 km. In addition, estimates were calculated for a modified Swarm trajectory that has Swarm A flying at a 50 km higher altitude (denoted Swarm mod.).}
    \label{fig:rcm-geometry}
\end{figure}

In the first part of this analysis, we use these virtual Swarm spacecraft trajectories from the period April-June 2014 to sample a single time-slice of the simulated magnetic perturbations and evaluate the corresponding curlometer estimates from three- and four-point estimates. We use these trajectories due to the close proximity and favourable configuration of the spacecraft in this period. To explore geometrical effects, we make a comparison to the results obtained for a regular tetrahedron of a comparable size constructed about the barycentre of the Swarm tetrahedron. Throughout this analysis, regular tetrahedron curlometer estimate is treated as the ground truth reference for comparisons of the current density. The simulation output contains the magnetic perturbations (obtained at LEO altitudes by Biot-Savart integration) and not the current densities directly. Therefore, local evaluation of the current density would require evaluating $\nabla \times \mathbf{B}$ using finite difference approximations on the geographic grid. However, it is not evident that such an approach would yield a more representative measure of the current density associated with the simulated magnetic perturbations than the curlometer. Moreover, the curlometer technique inherently estimates a volume-averaged current density over the spacecraft tetrahedron, whereas a finite-difference calculation would provide a local grid-based estimate. For these reasons, we adopt the curlometer estimate obtained from the regular tetrahedron as the reference measure for assessing geometric effects. To match the spatial and temporal resolution of the Swarm 1~Hz data, we interpolated the magnetic perturbations using the triangular-shaped cloud method \citep{Hockney_1988,Desai21b} to the location of the virtual spacecraft. To avoid grid effects, the interpolated perturbations were then low-pass filtered below the frequency corresponding to twice the grid spacing at the spacecraft velocity (about 120~km spatial scale). In the Swarm periods considered, Swarm B flies at an altitude slightly higher than the maximum simulated altitude. For this reason, we subtracted 50~km from the altitude of each Swarm spacecraft. This preserves the geometry of the tetrahedron without a large effect on the tetrahedron scale. 

The curlometer results for a particular Swarm trajectory are shown in Figure \ref{fig:rcm-geometry}. We find that the estimates calculated using the Swarm tetrahedron in configuration ABCCp produce large perpendicular currents similar to those shown in Figure \ref{fig:swarm1}. The estimates computed with the regular tetrahedron configuration show significantly smaller magnitudes of the perpendicular currents, which confirms that this erroneous amplification is a geometric effect of insufficient coverage in three dimensions by the Swarm tetrahedron faces. We note that the results for FACs between the different configurations (first panel) are in better agreement. Largest discrepancy is visible between 14:10 - 14:15 UT where the Swarm spacecraft cross over the pole, and the tetrahedron becomes highly distorted. For the regular tetrahedron, its volume is the same at every point of comparison, therefore this amplification does not occur in this case. Although highly planar, the orientation of the Swarm tetrahedron appears sufficiently well aligned with the field direction to produce reliable estimates of FACs, and ill-conditioning in perpendicular directions do not affect these. 

We additionally experimented with making small changes to the trajectories of Swarm spacecraft to generate a more regular configuration. The irregularity of the Swarm tetrahedron arises mostly due to the highly planar arrangement of the Swarm spacecraft, and its time-shifted counterparts (in this example delayed position of Swarm C). It should therefore be possible to improve the tetrahedron regularity by displacing one of the spacecraft out of the common orbital plane. The green-coloured lines in the top two plots of Figure \ref{fig:rcm-geometry} correspond to the Swarm configuration, with Swarm A flying at 50 km increased altitude (denoted Swarm mod.). Increasing Swarm A's altitude has the effect of improving the coverage of the tetrahedron faces in three dimensions, which is reflected in the Robert-Roux regularity parameter also shown in Figure \ref{fig:rcm-geometry}. The resulting increase in regularity has an apparent effect on the estimates for the perpendicular currents which are significantly reduced for the modified trajectory, while the overall agreement of FAC estimates is improved. To quantify the improvement, we can compare the root-mean-square difference, $\delta J_{(\parallel,\perp)}$, between the same-time estimates of the irregular (Swarm) configurations and the regular configuration for FACs and perpendicular currents averaged over the interval; For FACs, the modification of the Swarm tetrahedron leads to a decrease from $\delta J_\parallel$ = 0.15 \textmu A/m$^2$ to $\delta J_\parallel$ = 0.08 \textmu A/m$^2$ while for perpendicular currents the decrease is from $\delta J_\perp$ = 0.43 \textmu A/m$^2$ to $\delta J_\perp$ = 0.099 \textmu A/m$^2$. The trend between the three presented cases is also visible in the traces of $\text{div}\mathbf{B}/\text{curl}\mathbf{B}$; the ratio tends to be lower for configurations with higher regularity. However, we note that the present case study sampling of different volumes and different tetrahedron scales prevents from definitively isolating geometric effects on the accuracy of estimation. A clearer comparison could be achieved with Swarm tetrahedra of varying regularity flying through an analytical current structure, similar to the study conducted in \citet{Robert1998Accuracy} or \citet{FORSYTH2011598}. 
These results can inform the design of future low-Earth orbit multi-satellite constellations aimed at resolving FAC dynamics across different scales, defining the geometric requirements for reliable curlometer reconstruction and demonstrating how small orbit offsets can substantially improve current density retrievals.

\begin{figure}[h!]
    \centering
    \includegraphics[width=0.85\linewidth]{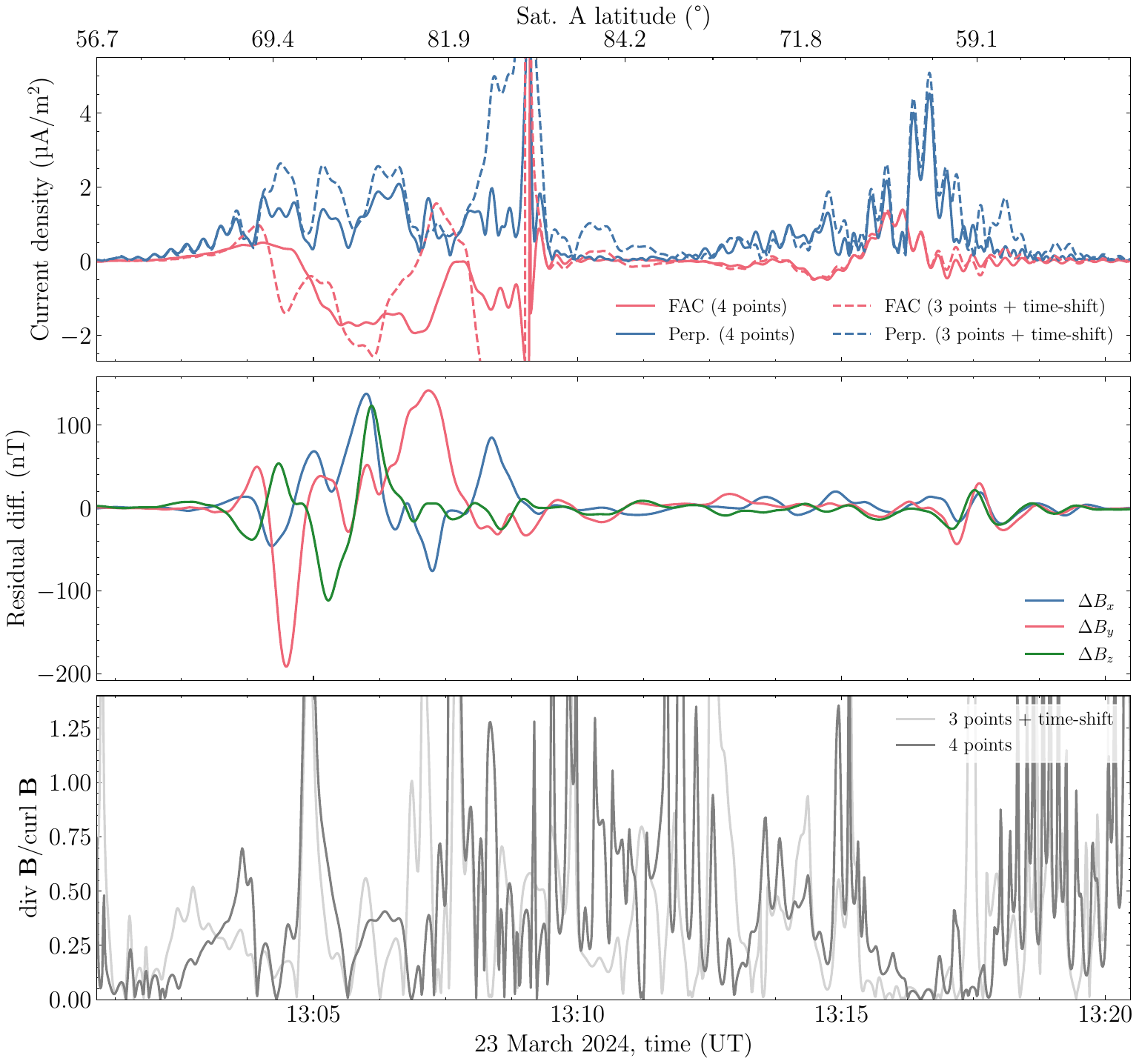}
\caption{Field-aligned and perpendicular components of the current density estimates (top) for the 29 May 2014, 13:26-13:46 UT, trajectory used to sample the temporally evolving magnetic perturbations (shown in Figure \ref{fig:polarB}), estimated using the curlometer with the Swarm tetrahedron in the configuration ABCCp with a 24 s time-shift (3-points + time-shift), and the same configuration with the time-shifted node Cp replaced by a physical node D (4 points). The configuration has a mean spacecraft separation between 120-250 km. The plot also shows the difference between the observed magnetic field perturbations (Cartesian components in SM) between the time-shifted node Cp and the 4th physical node D (middle) and the ratio of $\text{div}\mathbf{B}/\text{curl}\mathbf{B}$ (bottom).
}
    \label{fig:rcm_stationarity}
\end{figure}

\subsection{Time-stationarity}

In the second part of the analysis, we use the simulated magnetic perturbations evolving in time to check the effect of time-shifting on the current density estimates. Specifically, we compare the results between a virtual Swarm tetrahedron containing a time-shifted node, and the same tetrahedron with 4 physical nodes. The simulated time interval on the geographic grid together with the Swarm trajectory used for this analysis is shown in Figure \ref{fig:polarB}. We again use the configuration ABCCp for the time-shifted tetrahedron, and compare to the estimates obtained with the node Cp replaced by a physical node (denoted D). %

Figure \ref{fig:rcm_stationarity}, which plots the current density and difference in the magnetic perturbations observed by spacecraft Cp and D, shows that the spacecraft can observe significantly different fields that then lead to conflicting estimates of current densities. This is apparent in the first crossing of the auroral oval where differences of over 100~nT are apparent. This demonstrates that the time-shifted curlometer estimate often returns FACs orders of magnitude different to the ``true'' four point calculation. In the second crossing of the auroral oval, however, the estimates do not diverge significantly, indicating a highly-stationary FAC system across the scales of the time shift (24~s, 180~km). The $\text{div}\mathbf{B}/\text{curl}\mathbf{B}$ ratio in Figure \ref{fig:rcm_stationarity} displays a similar degree of asymmetry to the case of Swarm data shown in Figure \ref{fig:swarm1}, suggesting rapidly varying current structure on the side of first crossing of the auroral oval. This divergence during the first half of Figure 7, is also evident within Figure \ref{fig:polarB}, occurring on timescales of several minutes. The differences visible in Figure \ref{fig:rcm_stationarity} between current signatures around 14:05 UT  and 14:15 UT are likely caused by the rapid temporal evolution of the current system during this interval. Under disturbed geomagnetic conditions, FACs and associated magnetic perturbations can vary significantly on timescales of seconds due to transient Alfv\'en structures, filamentary current systems, and evolving auroral precipitation \citep{Gary1998,lysak2023kaw,zhou2025kilometer}. Consequently, the time-shifted virtual node may sample a substantially different magnetic structure from that observed by the physical fourth node, leading to discrepancies in the derived current density estimates. Furthermore, given the linear assumptions inherent in the curlometer technique, the divergence of $\text{div}\mathbf{B}/\text{curl}\mathbf{B}$ in the first part of the crossing could suggest that neither configuration shown in Figure \ref{fig:rcm_stationarity} provides an accurate estimate the current density during this interval, when the underlying current exhibits strong temporal or spatial variability on the tetrahedron scale (about 200 km in this case). Accurate estimation in this case would necessitate a configuration sampling the perturbations at smaller scales to characterise the small-scale variability. Future work should investigate how these differences depend on geomagnetic activity level, spatial scale, and tetrahedron geometry, and determine the limits under which the time-stationarity assumption remains valid.

\section{Discussion \& Conclusions}
In summary, we have analysed the capabilities and limitations of the curlometer method. Specifically, we have considered the impact of the time-stationarity assumption, needed to create the fourth measurement node for the application of the curlometer technique to Swarm spacecraft. In addition, we showed how limited intervals where the spacecraft constellation resembles a regular tetrahedron, necessary for reliable reconstruction of the current density in three dimensions, influence the estimates of FACs and perpendicular currents. 

Our results from correlation analysis indicate that time stationarity is often not met for a typical Swarm tetrahedron constructed with time-shifted data. Simulations further demonstrate that the time-shifted and the physical spacecraft may observe magnetic field amplitudes differing by $>10^{2}$ nT, resulting in substantial discrepancies between the estimated current densities.  

The degree to which a regular configuration can be constructed with time-shifted nodes is also limited. We have shown explicitly how poor geometry of the constellation can produce inaccuracies in the perpendicular currents, which are not sufficiently well covered by the tetrahedron faces. The results from the simulation also showed how small adjustments to the trajectories can significantly reduce those inaccuracies. This quantitatively determines it likely not possible to use Swarm to measure perpendicular currents in LEO, but future missions with gradiometric capabilties aligned with the field are well-suited for this.

A natural extension of this study would be to systematically investigate the validity of the time-stationarity assumption under both quiet and disturbed geomagnetic conditions. Here, all Swarm and simulated observations correspond to geomagnetically quiet intervals, and the use of several different Swarm periods showed that the main results were consistently reproduced across multiple encounters. The simulation interval was deliberately chosen independently of the observational intervals because the objective was to investigate the underlying curlometer technique rather than to reproduce the specific magnetic-field evolution observed by Swarm. Matching the simulation and observational intervals would instead introduce an additional dependence on their mutual temporal correlation, which is not relevant to the conclusions of this study. A systematic comparison of matched observations and simulations could nevertheless be used in future work to determine quantitative temporal windows over which the stationarity assumption remains valid as a function of geomagnetic activity, spatial scale, and orbital (tetrahedral) geometry. Across all observational and synthetic cases considered here, the unphysical perpendicular currents were consistently reproduced, while simple and realistic modifications to the Swarm orbits were shown to improve the tetrahedron geometry and substantially mitigate this limitation. Such an approach can inform the design of future gradiometric missions to LEO and potentially enable the first direct detection of perpendicular current closures at these altitudes.

The treatment of ill-conditioned matrices in linear inverse problems is well established and has been extensively developed in the literature \citep[e.g.][]{tikhonov1977solutions}. There exist methods which could be used to make the best use of available data even when the curlometer geometry may not be most suitable. A simple but useful diagnostic is the matrix condition number \citep{strang09}, which provides a quantitative measure of sensitivity to errors in the input data, analogous to how geometric quality factors are used. Such an approach could be determine intervals of estimates could then be rejected where the condition number exceeds a pre-defined threshold. However, rejection of poorly conditioned intervals is not always necessary. As demonstrated here, FAC estimates can remain robust even when the overall geometry is highly irregular, reflecting the fact that certain components of the current density vector may still be well constrained. Examining the matrix inversion terms could determine which directions are well-resolved. It may therefore be more appropriate to restrict the solution to a reduced-dimensional subspace, yielding reliable one- or two-dimensional current density estimates rather than a full three-dimensional reconstruction. There also exist well-established techniques such as singular value decomposition (SVD) or ridge regression \citep{vogel02,Vogt2020} which might be used to improve the  stability of estimates. 

These findings highlight the fundamental limitation of the time-shifted configurations and effects of irregular geometry, emphasising the importance of incorporating a fourth physical spacecraft in future missions to enable accurate gradiometric measurements without relying on stationarity assumptions.

\section*{Acknowledgements}
RTD acknowledges an STFC Ernest Rutherford Fellowship ST/W004801/1, award ST/Y005635/1, and ESA ROARS contract 4000142470/23/NL/AS/ov \citep{desai2026roars}. DL is supported by NSF Award AGS‐2347149, 80NSSC26K0295 and 80NSSC22M0064. MWD is supported by the NSFC Grant 42350710793 and Natural Environment Research Council (NERC) Grants NE/W003309/1 and NE/W002981/1. SRP is supported by JSPS Grant 22K21345. Contributions from MF and GH were supported by CNES through project ``Suivi et exploitation de la mission Swarm". MTW acknowledges funding from UKRI STFC through the Ernest Rutherford Fellowship (ST/X003663/1).

\section*{Data Availability Statement}
The Swarm satellite data used in this paper is openly available at \url{https://swarm-diss.eo.esa.int/}. Simulation data used in the analysis is openly available on Zenodo at \url{https://doi.org/10.5281/zenodo.22870976} and at the GitHub repository \url{https://github.com/rgski/curlometer}.

\bibliography{bibliography}
\end{document}